# Phase field modeling of crack propagation under combined shear and tensile loading with hybrid formulation

Heeyeong Jeong[a], Stefano Signetti[a], Tong-Seok Han[b], Seunghwa Ryu[a,*]

[a]Department of Mechanical Engineering & KI for the NanoCentury, Korea Advanced Institute of Science and Technology, Daejeon 34141, Republic of Korea

[b]Department of Civil and Environmental Engineering, Yonsei University, Seoul 03722, Republic of Korea

*corresponding author: ryush@kaist.ac.kr

## Abstract

The crack phase field model has been well established and validated for a variety of complex crack propagation patterns within a homogeneous medium under either tensile or shear loading. However, relatively less attention has been paid to crack propagation under combined tensile and shear loading or crack propagation within composite materials made of two constituents with very different elastic moduli. In this work, we compare crack propagation under such circumstances modelled by two representative formulations, anisotropic and hybrid formulations, which have distinct stiffness degradation schemes upon crack propagation. We demonstrate that the hybrid formulation is more adequate for modeling crack propagation problems under combined loading because the residual stiffness of the damaged zone in the anisotropic formulation may lead to spurious crack growth and altered load-displacement response.

# 1. Introduction

The initiation and propagation of cracks is one of the main failure mechanisms of engineering materials. Hence, numerous studies have been conducted to develop methods to accurately predict crack initiation and propagation under various mechanical loading conditions to prevent the catastrophic failure of engineering materials and systems. The theoretical basis to predict crack evolution was first introduced by Griffith [1] and Irwin [2]. They addressed the difficulty of dealing with a singular stress field at the crack tip by introducing the concept of the strain energy release, which showed good matching in numerous experiments on the initiation of pre-existing crack growth. However, crack nucleation, curvilinear crack paths, crack branching, or coalescence cannot be well accounted for.

In recent years, there has been an increasing interest in variational approaches to brittle fracture, which is referred to as the crack phase field model [3–7]. The phase field model approximates sharp crack discontinuity with a continuous scalar parameter denoted by the crack phase field. It has been shown that the solution of an approximated crack surface, described by a smooth function, converges to the solutions of sharp crack in the limit of regularization parameter equal to zero [8–13]. The phase field approach has attracted significant attention as a powerful tool to simulate complex crack evolution, including curvilinear crack paths, crack branching, or coalescence. Moreover, the phase field model framework has been extended beyond the linear elastic fracture regime to a wide range of fracture problems, such as large strain problems [14, 15], cohesive fractures [16], ductile fractures [17], multi-physics [18-23], pore microstructures[24], and dynamic effects [25, 26].

Although extensive phase field modeling studies have been carried out on the failure of homogeneous media, relatively less attention has been paid to the failure of composite materials, which inherently involves complex curvilinear crack propagation paths. In addition to accurately predicting curvilinear crack paths through composites, it is also crucial to obtain the entire stress–strain curve until complete failure to evaluate the toughness modulus of composites [27-29]. However, phase field methodologies [18, 19, 30, 31] based on the early formulation by Miehe [6,7] selectively degrade the stiffness along the direction of the maximum tensile strain upon crack growth. Hence, the cracked region unphysically sustains any subsequent loading with different maximum tensile direction. For example, a straight crack grown under tensile loading would withstand subsequent shear loading, as will be shown later. When the Young's moduli of two constituents in a composite are similar to each other, such behavior does not play a critical role in determining crack paths or predicting stress–strain curves because the crack paths are similar to those of homogeneous materials [31]. However, composites involving two constituents with highly different elastic moduli, such as natural or nature-inspired composites, fail due to the propagation of strongly curved cracks that are subject to highly varying combinations of tension and shear loading along the crack paths. In such circumstances, crack paths as well as the stress–strain curves are significantly affected by the aforementioned unphysical load bearing capacity of the cracked regions. For example, composite samples can sustain tensile loading although wavy cracks propagate through the entire sample dimension, as will be shown in this paper.

This paper compares the performance of two different formulations, the early anisotropic formulation by Miehe [6, 7] and a recent hybrid formulation [32] for modeling crack propagation in homogeneous materials under a sequence of different loading modes and for simulating strongly curved crack propagation in composite materials. The hybrid formulation, which was originally developed to reduce the computational cost [32], it is demonstrated to not suffer from the aforementioned unphysical load bearing capacity for the case studies considered here.

The remainder of this paper is organized as follows. In section 2, we review fundamental equations of the phase field approach to quasi-static brittle fracture and briefly introduce the numerical implementation scheme within the commercial software ABAQUS. For ease of comparison with previous studies, we have adopted the notations of Miehe et al. [6, 7]. Section 3 provides few modeling examples on crack propagation in homogeneous and heterogeneous media, which highlights the advantage of the hybrid formulation. In section 4, we summarize the paper and discuss directions for future research.

## 2. Methods

In this section, we briefly review diffusive crack representation in the phase field framework, and we introduce three different formulations, namely, isotropic, anisotropic, and hybrid schemes, that are classified according to the strain energy split and stiffness degradation scheme upon crack propagation. We then explain the numerical implementation in the commercial finite element software ABAQUS.

**2.1 Diffusive crack topology described by crack phase field**

Consider a domain $\Omega \subset \mathbb{R}^D$ and its boundary $\partial \Omega$ describing a cracked material in $D$ dimensional space (see Fig. 1). Let $\Gamma$ be a $D-1$ dimensional surface inside of domain $\Omega$. Here, $\Gamma$ represents the crack surface within the material. As depicted in Fig. 1a, the topology of a sharp crack can be described by the phase field scalar parameter $d(x) \in [0,1]$ with

$$d(x) = \begin{cases} 1, & \text{on } \Gamma \\ 0, & \text{otherwise} \end{cases}, \quad (1)$$

which represents the fully broken state of the material for $d = 1$ and the unbroken state of the material for $d = 0$ at a given point $x$. In the regularized framework shown in Fig. 1b, the crack topology is approximated by scalar parameter $d(x)$ having a unit value on the crack surface $\Gamma$ and fading away from that surface. The value of the phase field $d(x)$ can be determined by solving the following differential equation:

$$\begin{cases} d - l^2\nabla^2 d = 0, & \text{in } \Omega \\ d(x) = 1, & \text{on } \Gamma \\ \nabla d(x) \cdot \mathbf{n} = 0, & \text{on } \partial\Omega \end{cases}, \tag{2}$$

where $\nabla^2 d$ is the Laplacian of the phase field, $\mathbf{n}$ is the outward normal on $\partial\Omega$, and $l$ is the regularization parameter that determines the width of the regularized or diffusive crack topology. To elaborate the concept of the regularization parameter $l$, a one-dimensional example of a diffusive crack for various values of $l$ is shown in Fig. 2b. In the limit of $l \to 0$, Fig. 2b shows that the diffusive crack topology converges to the ideal sharp crack. Similarly, in two-dimensional and three-dimensional cases, the diffusive crack topology also converges to a sharp crack for vanishing value of $l$. Diffusive crack topology $\Gamma_l(d)$ can be expressed as

$$\Gamma_l(d) = \int_\Omega \gamma(d, \nabla d)\, dV, \tag{3}$$

where $\gamma(d, \nabla d)$ is the crack surface density function per unit volume of the material, denoted as

$$\gamma(d, \nabla d) = \frac{1}{2l}d^2 + \frac{l}{2}|\nabla d|^2. \tag{4}$$

In terms of $\gamma(d, \nabla d)$ and the critical energy release rate $g_C$, we can approximate the surface energy $W(d)$ by volume integral as

$$W(d) = \int_\Gamma g_C\, dA \approx \int_\Omega g_C \gamma(d, \nabla d)\, dV. \tag{5}$$

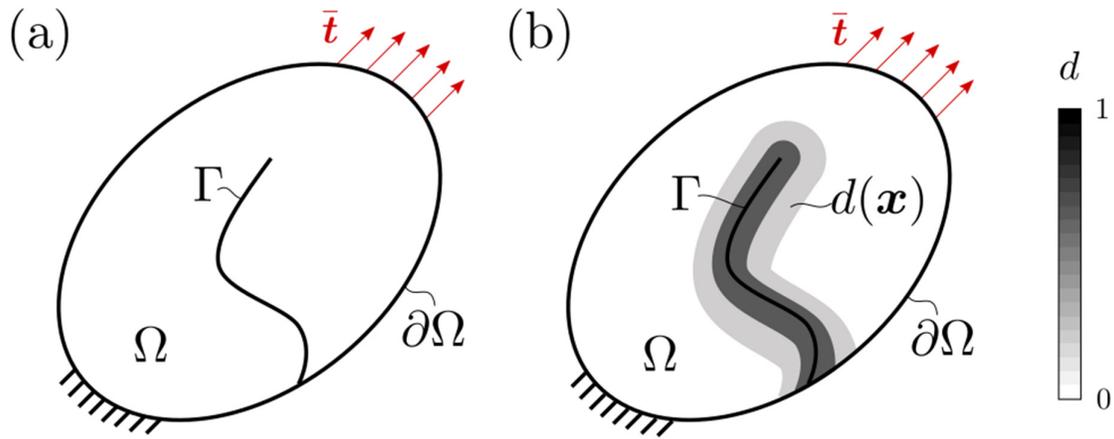

**Fig. 1.** Two-dimensional crack topology: (a) sharp crack model, (b) diffusive crack model described by phase field function $d(x)$.

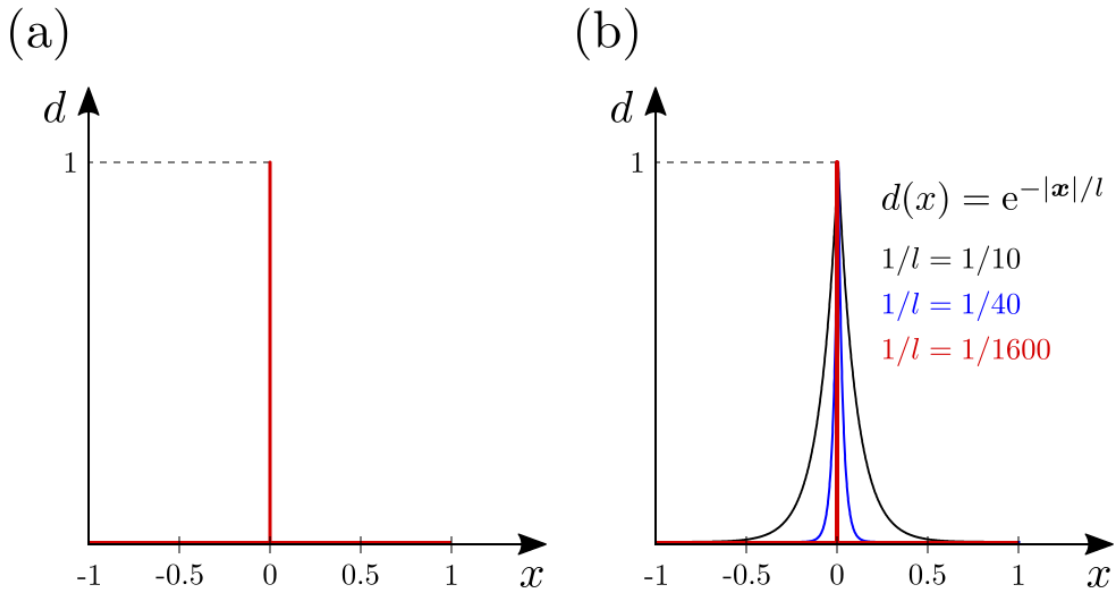

**Fig. 2.** Representation of one-dimensional crack at $x = 0$: (a) sharp crack model, (b) diffusive crack model described by phase field $d(x) = e^{-|x|/l}$ for different values of the regularization parameter $l$.

**2.2 Strain energy and stiffness degradation of fracturing material**

When the strain energy stored at a point of the material exceeds the energy required to open a crack surface, fracture starts and it is accompanied by both strain energy and stiffness degradation. In other words, the crack phase field $d(x)$ is driven by the strain energy of the material, and the completely fractured region with $d(x) = 1$ no longer sustains the mechanical loading. To couple the crack phase field $d(x)$ with displacement field $u(x)$, we define the strain energy of a material $E(u, d)$ as

$$E(u, d) = \int_\Omega \psi(\varepsilon(u), d) dV, \tag{6}$$

where $\varepsilon(u)$ is the strain tensor, and $\psi(\varepsilon(u), d)$ is the strain energy stored per unit volume of the material. Here, the value of $\psi$ depends not only on the displacement $u(x)$ but also on the crack phase field $d(x)$. Now, we turn our attention to the constitutive assumptions concerning the degradation of strain energy and stiffness that are directly related to the driving force of crack propagation. Depending on the constitutive assumptions regarding strain energy degradation, there are two major formulations, namely, isotropic and anisotropic. More detailed discussions can be found in the works of Miehe et al. [6, 7]. We also introduce a new hybrid formulation that shares the same strain energy degradation with the anisotropic formulation but an identical stiffness degradation scheme with the isotropic formulation, which was originally introduced to reduce the computational cost [32]. Fig. 3 provides a visual guide to the boundary conditions and variables in the governing equations presented below. We note that the shear friction along the crack path is not taken into account in the following formulations.

- Isotropic formulation:

$$\psi(\varepsilon(u), d) = (1-d)^2 \psi_0(\varepsilon) \tag{7}$$

$$\begin{cases} \sigma(u, d) = (1-d)^2 \dfrac{\partial \psi_0(\varepsilon)}{\partial \varepsilon} \\ -l^2 \Delta d + d = \dfrac{2l}{g_C}(1-d)H \end{cases} \tag{8}$$

Note that $\psi_0(\varepsilon) = \lambda \mathrm{tr}^2[\varepsilon]/2 + \mu \mathrm{tr}[\varepsilon^2]$ is the standard strain energy of an undamaged elastic material with Lame's parameters $\lambda$, $\mu$, and $\boldsymbol{\sigma}(\boldsymbol{u}, d)$ is the stress tensor as a function of the displacement field and crack phase field. Here, $H = \max_{\tau \in [0,t]} \psi_0(\boldsymbol{\varepsilon}(\boldsymbol{x}, \tau))$ is the so-called history variable, which ensures the irreversibility condition to prevent crack healing. Still, the isotropic model shows physically unrealistic crack evolution because the formulation allows for cracking in both compression and tension. Thus, in general, the isotropic formulation is applicable only to mode-I fracture.

- Anisotropic formulation:

$$\psi(\boldsymbol{\varepsilon}(\boldsymbol{u}), d) = (1-d)^2 \psi_0^+(\boldsymbol{\varepsilon}) + \psi_0^-(\boldsymbol{\varepsilon}) \tag{9}$$

$$\begin{cases} \boldsymbol{\sigma}(\boldsymbol{u}, d) = (1-d)^2 \dfrac{\partial \psi_0^+(\boldsymbol{\varepsilon})}{\partial \boldsymbol{\varepsilon}} + \dfrac{\partial \psi_0^-(\boldsymbol{\varepsilon})}{\partial \boldsymbol{\varepsilon}} \\ -l^2 \Delta d + d = \dfrac{2l}{g_C}(1-d)H^+ \end{cases} \tag{10}$$

Inspired by the earlier work of Amor et al. [5] which introduced an additive decomposition of the elastic energy density into volumetric and deviatoric contributions to distinguish between fracture behaviors in tension and compression, the anisotropic formulation was proposed to overcome the drawbacks of the isotropic formulation operating a tension-compression splitting by decomposing strain tensor $\boldsymbol{\varepsilon} = \sum_i^D \varepsilon^i \boldsymbol{n}^i \otimes \boldsymbol{n}^i$ into positive parts $\boldsymbol{\varepsilon}_+$ and negative parts $\boldsymbol{\varepsilon}_-$ [6, 7]:

$$\begin{cases} \boldsymbol{\varepsilon} = \boldsymbol{\varepsilon}_+ + \boldsymbol{\varepsilon}_- \\ \boldsymbol{\varepsilon}_+ = \sum_i^D \langle \varepsilon^i \rangle_+ \boldsymbol{n}^i \otimes \boldsymbol{n}^i \\ \boldsymbol{\varepsilon}_- = \sum_i^D \langle \varepsilon^i \rangle_- \boldsymbol{n}^i \otimes \boldsymbol{n}^i \end{cases} \tag{11}$$

Here, $\{\varepsilon^i\}_{i=[1,D]}$ denotes the principal strains, and $\{\boldsymbol{n}^i\}_{i=[1,D]}$ denotes the unit vector along the direction of principal strain. Bracket operators are defined as $\langle \bullet \rangle_+ = (\bullet + |\bullet|)/2$ and $\langle \bullet \rangle_- = (\bullet - |\bullet|)/2$. With the decomposed strain tensor, the positive part of the strain energy

$\psi_0^+(\varepsilon)$ and the negative part of the strain energy $\psi_0^-(\varepsilon)$ can be defined as

$$\psi_0^+(\varepsilon) = \lambda \langle \text{tr}[\varepsilon] \rangle_+^2/2 + \mu \text{tr}[\varepsilon_+^2], \qquad (12.\text{a})$$

$$\psi_0^-(\varepsilon) = \lambda \langle \text{tr}[\varepsilon] \rangle_-^2/2 + \mu \text{tr}[\varepsilon_-^2]. \qquad (12.\text{b})$$

In the anisotropic formulation, only the tensile part of the strain energy drives the crack evolution. Similar to the isotropic formulation, irreversibility condition is ensured by history variable $H^+ = \max_{\tau \in [0,t]} \psi_0^+(\varepsilon(x,\tau))$. Although the anisotropic formulation overcomes the limitations of the isotropic formulation, it still has a serious drawback. Since stiffness degradation happens only in the direction orthogonal to the crack path, even the fully broken region of the material can still support the mechanical load in the other directions. Such degradation in strain energy may cause physically unrealistic fracture patterns and post-critical behavior for various cases involving combined shear and tension.

- Hybrid formulation:

$$\psi(\varepsilon(u), d) = (1-d)^2 \psi_0^+(\varepsilon) + \psi_0^-(\varepsilon) \qquad (13)$$

$$\begin{cases} \sigma(u,d) = (1-d)^2 \dfrac{\partial \psi_0(\varepsilon)}{\partial \varepsilon} \\ -l^2 \Delta d + d = \dfrac{2l}{g_C}(1-d)H^+ \\ \forall x: \ \psi_0^+ < \psi_0^- \rightarrow d \equiv 0 \end{cases} \qquad (14)$$

To overcome the limitation of the anisotropic formulation, we propose to exploit the hybrid formulation originally proposed by Ambadi et al. [32] to reduce the computational cost. We find that crack evolution and strain energy degradation can be modeled more accurately with the hybrid formulation. As in the anisotropic formulation, crack evolution is driven only by the tensile part of the strain energy, which in turn prevents crack evolution under compression. However, unlike the case of anisotropic formulation, stiffness degradation happens in all directions of the material. Consequently, the fully broken part of the material cannot support the load in any direction, as will be shown for a variety of examples in Section 3. The last constraint in Eq. (14) is adopted to prevent crack face inter-penetration under compression [32] that results from complete stiffness degradation. Therefore, the hybrid formulation predicts physically appropriate failure mechanisms when we consider samples subjected to combined tension and shear loading. For all three models, the governing quasi-static equilibrium equation for a displacement field with appropriate traction and displacement boundary conditions should satisfy the following:

$$\begin{cases} \text{div}[\boldsymbol{\sigma}(\boldsymbol{u},d)] + \boldsymbol{b} = 0, & \text{in } \Omega \\ \boldsymbol{\sigma} \cdot \boldsymbol{n} = \bar{\boldsymbol{t}}, & \text{on } \partial\Omega_\sigma \\ \boldsymbol{u} = \bar{\boldsymbol{u}}, & \text{on } \partial\Omega_u, \\ d = 1 & \text{on } \Gamma \\ \nabla d \cdot \boldsymbol{n} = 0 & \text{on } \partial\Omega \end{cases} \quad (15)$$

where $\boldsymbol{b}$ is the body force; $\boldsymbol{n}$ is the normal vector specified for the traction boundary; and $\bar{\boldsymbol{t}}$ and $\bar{\boldsymbol{u}}$ are the prescribed traction and displacement at the loaded and constrained boundaries, respectively (Figure 3). We note that the Neumann-type boundary condition on the phase field parameter allows $d \neq 0$ when a crack propagates toward the boundary without imposing an unphysical constraint, and at the same time is compatible with $d = 0$ at the boundary when a regularized crack is sufficiently far from the boundary.

### 2.3. Numerical implementation

We numerically solved the phase field models based on the finite-element method following the approaches of Miehe et al. [7]. A staggered scheme, in which the displacement field and the crack phase field are updated alternatively, was chosen to ensure the robustness of the numerical solutions. Additionally, for further numerical robustness and to lower the computational cost, we performed decomposition of the strain tensor in a previous loading step based on the shifted strain tensor split algorithm [31], which is expressed as

$$\boldsymbol{\varepsilon}_{n+1}^+ = \boldsymbol{P}_n^+ : \boldsymbol{\varepsilon}_{n+1}, \qquad (16.a)$$

$$\boldsymbol{\varepsilon}_{n+1}^- = \boldsymbol{P}_n^- : \boldsymbol{\varepsilon}_{n+1}. \qquad (16.b)$$

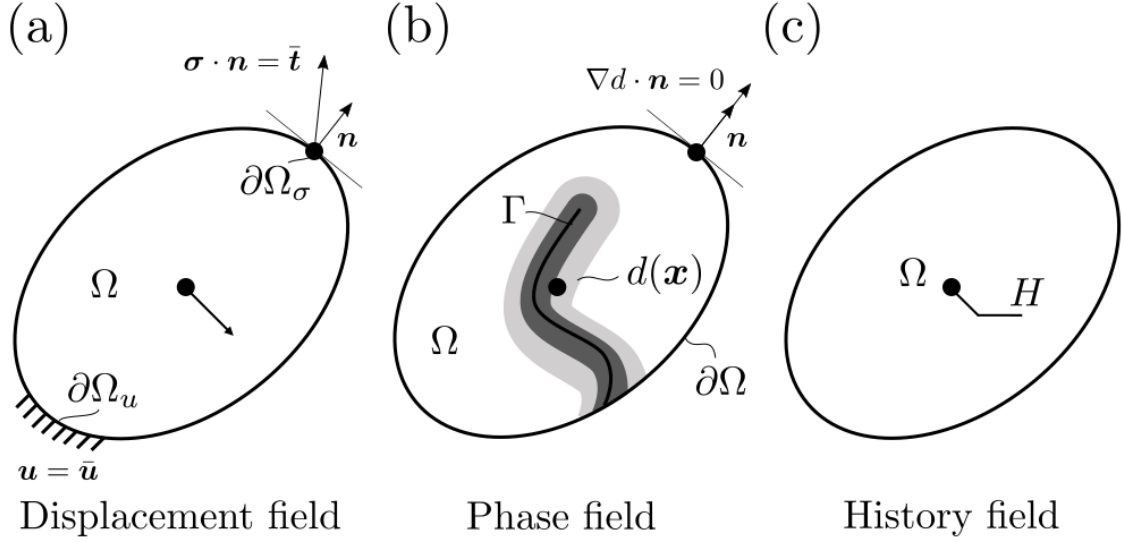

**Fig. 3.** Multi-field approach of crack propagation problem in elastic solid. (a) The displacement field is subject to the Dirichlet-type boundary condition $\boldsymbol{u} = \bar{\boldsymbol{u}}$ on $\partial\Omega_u$ and the Neumann-type boundary condition $\boldsymbol{\sigma} \cdot \boldsymbol{n} = \bar{\boldsymbol{t}}$ on $\partial\Omega_\sigma$. (b) The phase field is subject to the Dirichlet-type boundary condition $d = 1$ on $\Gamma$ and the Neumann-type boundary condition $\nabla d \cdot \boldsymbol{n} = 0$ on $\partial\Omega$. (c) The history field is defined by a maximum local strain energy through the whole fracture process.

Here, $\boldsymbol{\varepsilon}_n$ and $\boldsymbol{\varepsilon}_{n+1}$ are strain tensors at times $t_n$ and $t_{n+1}$, respectively. Also, $\boldsymbol{P}_n^\pm = \frac{\partial \boldsymbol{\varepsilon}_n^\pm}{\partial \boldsymbol{\varepsilon}}$ denotes a fourth-order tensor whose columns are composed of the eigenvectors of the strain tensor $\boldsymbol{\varepsilon}_n$.

We implemented the crack phase field solver for the quasi-static fracture of elastic solid in the commercial software ABAQUS, by adopting the procedures described in [33-35]. We were able to take advantage of the efficiency of ABAQUS in solving nonlinear problems. Using its user-defined element (UEL) subroutine, two types of quadrilateral four-node elements in 2D (plane stress) were constructed in a layered manner. A schematic illustration of layered element structures in ABAQUS [36] is shown in Fig. 4. The UEL subroutine in ABAQUS allows users to define element stiffness matrices as well as residual load vectors. The constitutive behavior of each element is determined by calling the UEL subroutine. The first and second type elements contribute to the phase field and displacement, respectively. Displacement $\boldsymbol{u}$ and phase field $d$ are discretized with node values $\boldsymbol{u}_i$ and $d_i$ as:

$$\boldsymbol{u} = \sum_{i=1}^{4} N_i \boldsymbol{u}_i, \tag{17}$$

$$,d = \sum_{i=1}^{4} N_i d_i \tag{18}$$

where $N_i$ denotes the shape function correspond to node $i$. Also, we can express the strain tensor and the gradient of the phase field as

$$\boldsymbol{\varepsilon} = \sum_{i=1}^{4} \boldsymbol{B}_i^u \boldsymbol{u}_i, \tag{19}$$

$$\nabla d = \sum_{i=1}^{4} \boldsymbol{B}_i^d d_i, \tag{20}$$

where $\boldsymbol{B}_i^u$ and $\boldsymbol{B}_i^d$ are the corresponding matrices of spatial derivatives of the shape functions given as

$$\boldsymbol{B}_i^u = \begin{bmatrix} N_{i,x} & 0 \\ 0 & N_{i,y} \\ N_{i,y} & N_{i,x} \end{bmatrix}, \tag{21.a}$$

$$\boldsymbol{B}_i^d = \begin{bmatrix} N_{i,x} \\ N_{i,y} \end{bmatrix}. \tag{21.b}$$

To calculate the value of $\boldsymbol{u}$ and $d$ at each load step, the following equation system is solved iteratively with the quasi-Newton method:

$$\begin{bmatrix} \boldsymbol{K}_n^d & 0 \\ 0 & \boldsymbol{K}_n^u \end{bmatrix} \begin{bmatrix} d_{n+1} \\ \boldsymbol{u}_{n+1} \end{bmatrix} = - \begin{bmatrix} r_n^d \\ r_n^u \end{bmatrix}, \tag{22}$$

where $\boldsymbol{K}_n^d$ and $\boldsymbol{K}_n^u$ are the stiffness matrices of phase field and displacement, respectively, $d_{n+1}$ and $\boldsymbol{u}_{n+1}$ are the unknown vectors of phase field and displacement at each node with $r_n^d$ and $r_n^u$ being the residual load vectors associated with the crack phase field and the displacement field, respectively. The corresponding matrices and vectors are evaluated by the following equations:

$$K_{ij}^d = \int_\Omega [B_i^d]^T g_C l [B_j^d] + N_i \left(2H^+ + \frac{g_C}{l}\right) N_j \, dV, \tag{23}$$

$$K_{ij}^u = \int_\Omega [B_i^u]^T \frac{\partial \sigma}{\partial \varepsilon} [B_j^u] \, dV, \tag{24}$$

$$\frac{\partial \sigma}{\partial \varepsilon} = \frac{\partial}{\partial \varepsilon}\left(\frac{[1-d]^2 \partial \psi_0^+}{\partial \varepsilon} + \frac{\partial \psi_0^-}{\partial \varepsilon}\right), \tag{25}$$

$$r_n^d = \int_\Omega g_C \left(l[B_i^d]^T \nabla d + \frac{1}{l} N_i d\right) - 2(1-d)N_i H^+ \, dV, \tag{26}$$

$$r_n^u = \int_\Omega [B_i^u]^T \sigma \, dV - \int_{\partial \Omega_\sigma} N_i \bar{t} \, dA \tag{27}$$

Because the shape functions of the element are user-defined in the UEL subroutine, the ABAQUS post processing software could not automatically extrapolate variables at each integration point to element nodes, which would require users to make an extra effort for the post-processing. Therefore, by user-defined material (UMAT) subroutine we added a third layer for the visualization of the calculated crack phase and displacement fields. With the UMAT subroutine, a fictitious mesh consisting of ABAQUS native elements was overlaid. By matching the shape function for the ABAQUS native elements with the shape function of the UEL subroutine, we were able to visualize our results from the UEL subroutine. Information of the UEL subroutine was transferred to the UMAT subroutine using "COMMON" statement.

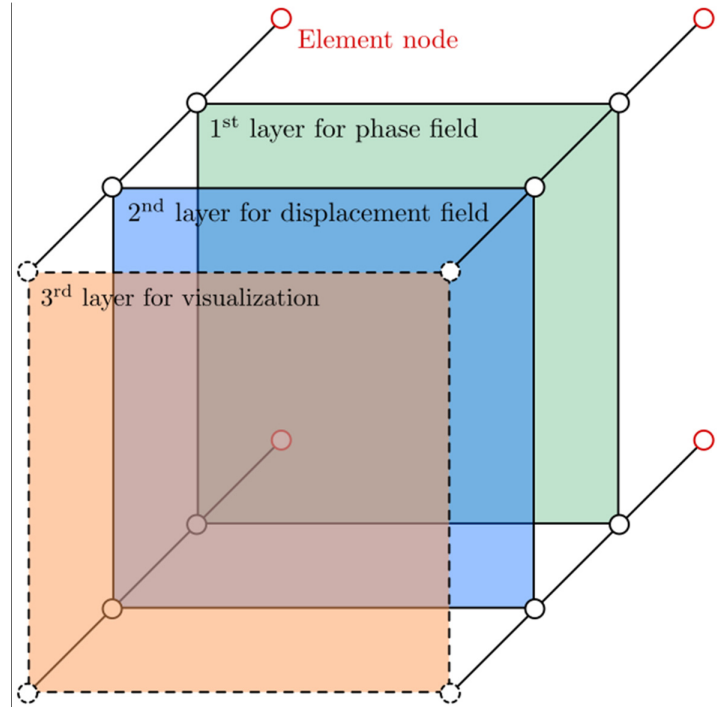

**Fig. 4.** The layered element structure in ABAQUS developed in this study. All three layers share the same nodes and just 1$^{st}$ and 2$^{nd}$ ones contribute to the stiffness of different degrees of freedom.

## 3. Numerical examples

In this section, we show four numerical examples that highlight the differences between the hybrid and the anisotropic formulations. These examples show that the hybrid formulation leads to physically adequate results. We consider two single-notched homogeneous specimens. The first one is subject to mode-I and mode-II loading in a sequential manner, while the second one is subject to combined tension and shear loading. We then consider two composite specimens made of two constituent phases with extreme elastic modulus mismatch. The first one is a bone-inspired composite with spherical embedding inclusions, and the second one is a nacre-inspired composite with a rectangular high-aspect-ratio embedding platelets. All constituent materials in this work follow a linear elastic and isotropic constitutive law.

## 3.1 Fracture of single-edge notched specimen under sequential tension and shear

We demonstrate the limitation of the anisotropic formulation (Eqs (9)-(12)) by simulating the fracture of a homogeneous material under a sequential application of tensile and shear loadings. The geometry and the boundary conditions are depicted in Fig. 5a. We set the materials parameters to the following values: Young's modulus $E = 210$ kN/mm$^2$, Poisson's ratio $\nu = 0.3$, $l = 0.01$ mm, and $g_C = 5 \cdot 10^{-4}$ kN/mm. The specimen is modelled with a structured mesh of 40,000 quadrilateral elements with grid size of 0.005 mm. Fig. 5b depicts the displacement loading condition imposed on the top edge of the specimen. Tensile loading was first applied until a crack propagated through the entire specimen, which then was followed by shear loading.

As plotted in Fig. 5c, the results obtained by the two formulations were identical during the tensile loading stage. In contrast, upon the subsequent shear loading, the anisotropic formulation and the hybrid formulation resulted in completely different load-displacement curves, as shown in Fig. 5d. Figs. 6 show the crack patterns at the end of the shear loading for the anisotropic and hybrid formulations, respectively. With the anisotropic formulation, the completely fractured material after tension still withstood shear loading, resulting in shear deformation. In contrast, the shear force remained at a zero value in the case of the hybrid formulation. Accordingly, the two fractured crack surfaces, formed during the first tensile loading, slide without shape change when modeled by the hybrid formulation. This is a direct consequence of the anisotropic formulation because stiffness degradation occurs only in the direction of maximum tensile loading, while a finite stiffness remains along the sliding direction.

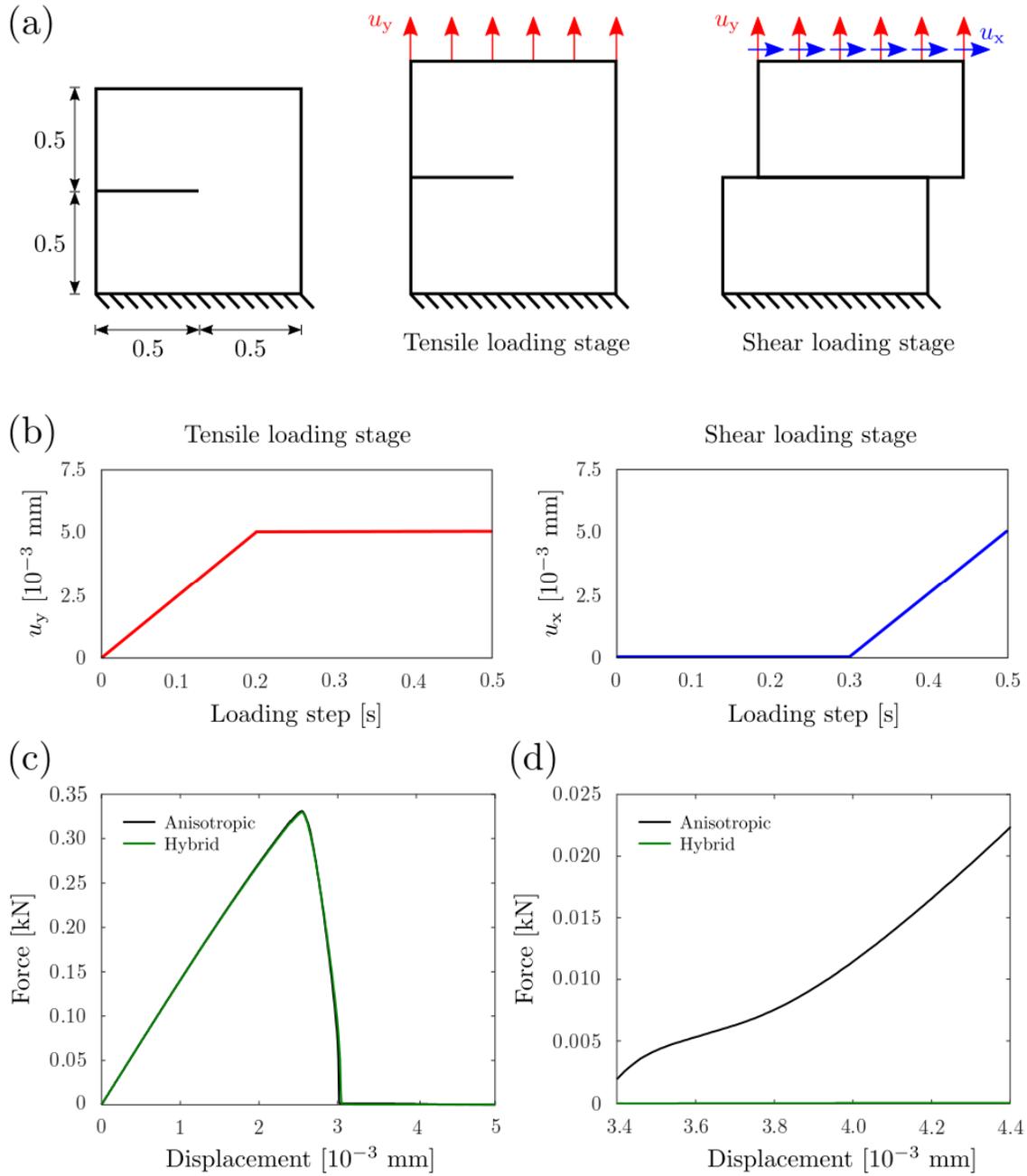

**Fig. 5.** (a) Geometry and boundary condition for the homogeneous single-edge notched specimen (units of measure are mm). (b) Loading history for tensile and shear loads applied in series. Resulting force-displacement curves for (c) tensile loading stage and (d) shear loading stage.

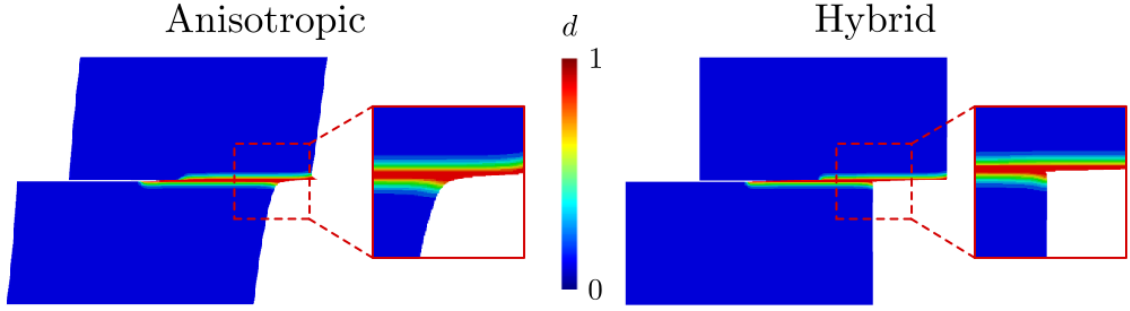

**Fig. 6.** Visualization of the phase field within the deformed specimen after tension and shear for anisotropic and hybrid formulations (displacement along *x* direction is artificially multiplied by a magnifying factor).

**3.2 Bending of a homogenous beam with fully clamped ends**

In this example, we chose a single-notched beam with fixed ends undergoing bending under indentation. Indentation was applied at two different locations along the beam to compare the outcomes of the anisotropic formulation and the hybrid formulation. In case I, loading was applied at the center of the beam (Fig. 7a), and the crack growth was driven by mode I fracture. In case II, loading was eccentric (Fig. 8a) so that crack propagation was enabled by combined mode I and mode II fracture. We set the material parameters as follows: $E = 20.8 \text{ kN/mm}^2$, $\nu = 0.3$, $l = 0.0125$ mm, and $g_C = 5 \cdot 10^{-4}$ kN/mm. In the central part of the beam (within 0.5 mm from crack location) the average grid size is of $\sim 0.006$ mm and 0.05 mm elsewhere, with a total number of elements of 37,479. The initial crack length was set as 0.2 mm (Fig. 7a).

The force-displacement curves and crack patterns simulated by the hybrid and anisotropic formulations for case I are shown in Fig. 7b-d. Both formulations lead to a qualitatively similar behavior; the crack growth initiates at a critical load, and crack propagation is saturated when the crack length reaches about 0.8 mm due to the reduction of the energy release rate due to the additional level of restrain at the fully clamped ends. This is different from the typical fracture behavior of the beam with free ends under 3-point bending where catastrophic crack growth occurs beyond a critical indentation force. Case I modeling confirms that, when the crack growth is driven only by mode I fracture, the two formulations result in almost indistinguishable crack growth patterns and force-displacement curves.

In contrast, the force-displacement and crack patterns resulting from the two formulations are significantly different, as depicted in Fig. 8b,c. In both formulations, cracks propagated in a curvilinear way because the crack tip is subjected to the combined tension and shear loading when indentation is not applied at the center of the beam. However, cracks propagate further when the hybrid formulation is used. The force-displacement curves exhibit similar behavior in the initial stage, but they start to deviate as the cracks start to propagate; the anisotropic formulation predicts a larger load in comparison to the hybrid formulation. Because the strain energy degrades only along the normal direction of the crack path in the anisotropic formulation, the cracked region sustains extra shearing load, which increases the resistance to crack growth. It is evident from this case how the anisotropic formulation may provide significant overestimation of the bearing capacity of cracked members under mixed-mode stress field.

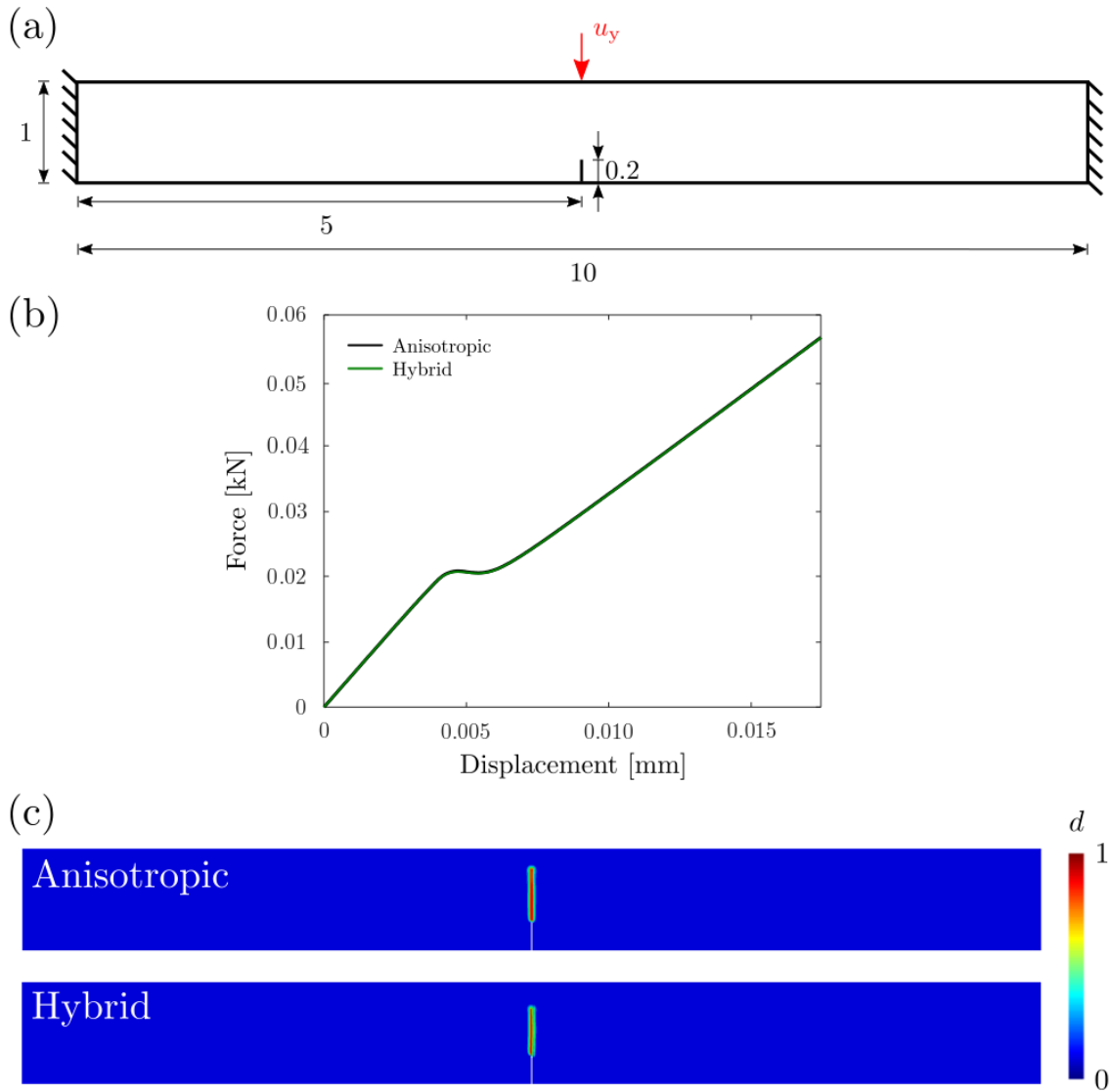

**Fig. 7.** (a) Geometry and boundary condition for single-edge notched fully clamped beam indented at the midspan (case I, units of measure are mm). (b) Resulting force-displacement curves. (c) Crack patterns for anisotropic and hybrid formulations with contour plot of the phase field parameter.

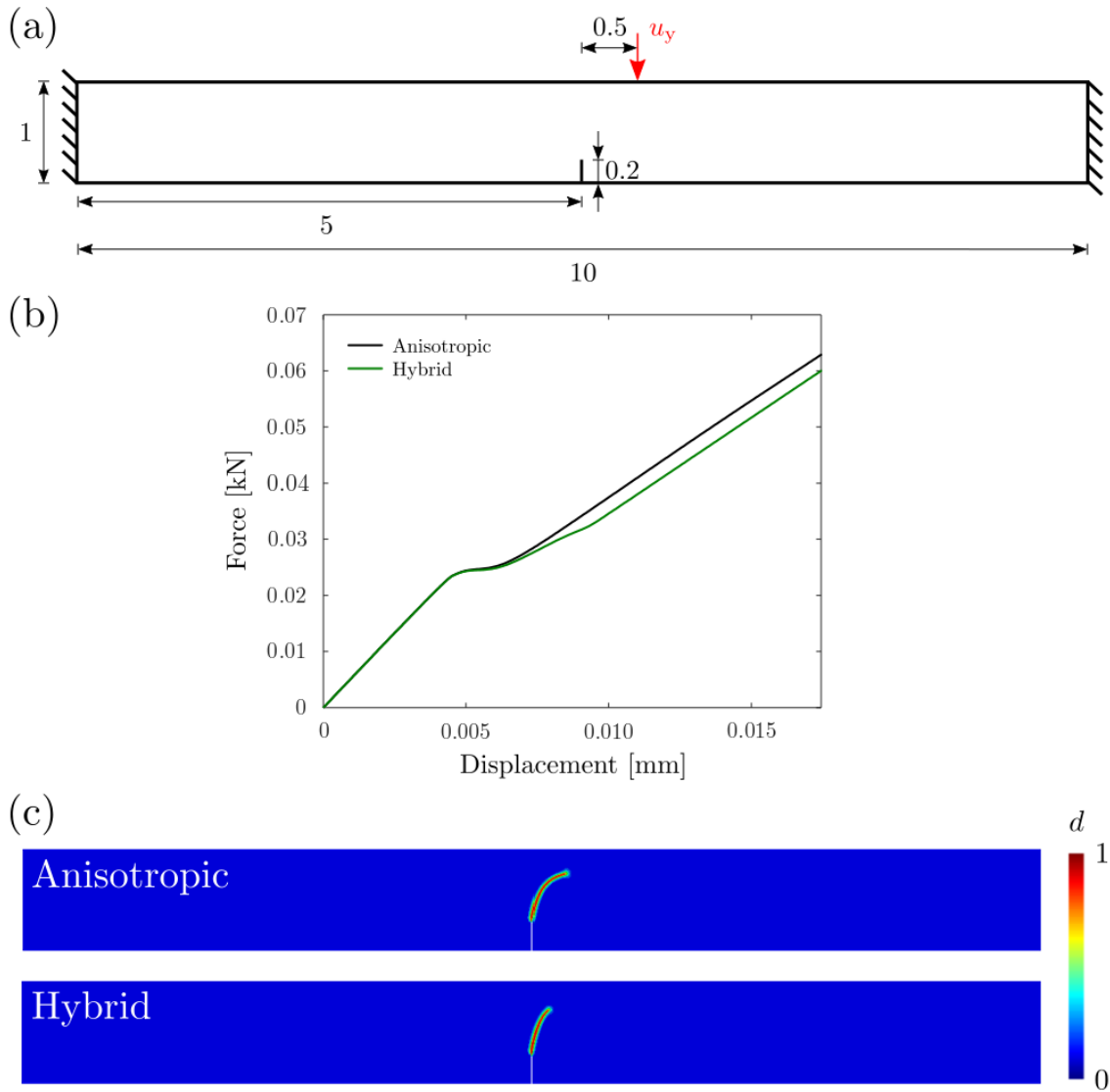

**Fig. 8.** (a) Geometry and boundary condition for single-edge notched fully clamped beam with non-centered indentation load (case II, units of measure are mm). (b) Resulting force-displacement curves. (c) Crack patterns for anisotropic and hybrid formulations with contour plot of the phase field parameter.

## 3.3 Tensile test of a bone-inspired composite

Having tested the crack propagation in homogeneous specimens, we now consider the fracture of heterogeneous materials composed of two constituent materials with very high mismatch in the elastic moduli, such as those occurring in biomaterials or engineering composites reinforced with high strength and high modulus fibers. We first consider a specimen designed to mimic the microstructure of cortical bone, following previous studies [27, 28]. It is known that a high modulus mismatch in cortical bone structure causes crack deflection [27, 28], which increases the toughness modulus of the material. For the design of composites inspired by such structures and for the optimization of their toughness, it is crucial to predict the entire force-displacement curve accurately. We use a single-notched specimen for the case study. The modulus of the hard material (inclusion) was set to be 100 times larger than that of the soft material. The geometry and the boundary condition are shown in Fig. 9a. The simulation parameters were set as follows: $E_{\text{hard}} = 2100$ kN/mm$^2$, $E_{\text{soft}} = 21$ kN/mm$^2$, $\nu = 0.3$, $l = 0.006$ mm, and $g_C = 5 \cdot 10^{-5}$ kN/mm. Here, the Poisson's ratios, regularization parameters, and the fracture energies were assumed the same for both phases. The total number of elements was 148,921 with average grid size of 0.003 mm.

The crack patterns for each loading step for two formulations are shown in Fig. 9d. In the initial stage of crack growth, the crack patterns from the two formulations are not significantly different to each other, but they start to deviate after the crack has completely propagated to the whole specimen. In the case of anisotropic formulation, the crack keeps spreading into the soft matrix even after the complete propagation of the primary crack (see Fig. 9d). In addition, as plotted in Fig. 9c, the load acting on the specimen does not reduce to zero, as expected, even showing an unphysical hardening behavior. Again, this occurs because the wavy crack path is subjected to the combined tension and shear stresses, although the composite specimen is subjected to the pure tension. As expected, the simulation based on the hybrid formulation does not suffer from such problem.

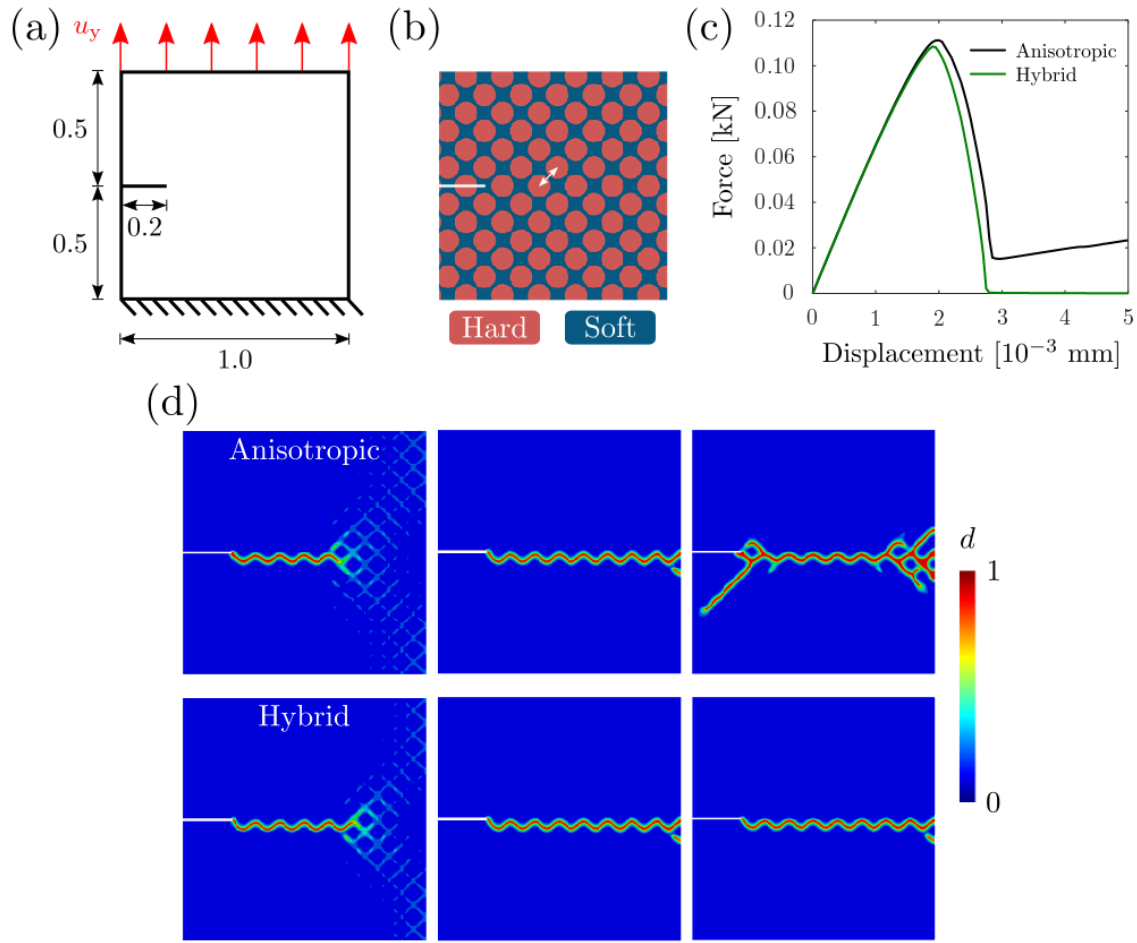

**Fig. 9.** (a) Geometry and boundary condition for single-edge notched specimen (units of measure are mm). (b) Composition of the composite specimen with a bone-inspired structure (radius of the circular inclusion 0.064 mm, centre-to-centre distance, specified with a double arrow, 0.142 mm). (c) Force-displacement curves for the tension test. (d) Crack patterns for anisotropic and hybrid formulations at three different loading steps ($u_y = 2.5 \times 10^{-3}$ mm, $u_y = 5.0 \times 10^{-3}$ mm, $u_y = 12.5 \times 10^{-3}$ mm).

## 3.4 Tension test on a nacre-inspired composite

In this section, we present the testing of a specimen mimicking the "brick and mortar" microstructure of nacre that fails by even more severely curved cracking [29]. As in the bone-inspired composite, we set the modulus of the hard material to be 100 times larger than that of a soft material, and we tested a single-notched specimen. The geometry and the boundary conditions are shown in Fig. 10a. The simulation parameters were set as follows: $E_{\text{hard}} = 2100$ kN/mm², $E_{\text{soft}} = 21$ kN/mm², $\nu = 0.3$, $l = 0.004$ mm, and $g_C = 5 \cdot 10^{-5}$ kN/mm. Again, the Poisson's ratios, regularization parameters, and the fracture energies were assumed the same for both phases. The total number of elements was $258{,}749$ with an average grid size of $0.002$ mm. The force-displacement curves for both formulations are plotted in Fig. 10c, and the crack patterns for each loading step are presented in Fig. 10d.

Interestingly, the crack patterns of the nacre-like composite with the two different formulations significantly differed. The crack modelled by the hybrid formulation grows along the soft mortar region and the load-displacement curve approaches zero as the crack propagates through the entire sample. In contrast, the crack simulated by the anisotropic formulation grows toward one boundary, then deflects towards the opposite boundary, and deflects again until the crack propagate through the whole span of the sample. Because the crack path can withstand shear loading, the force-displacement curve never converges to zero, and the crack becomes wider with time. This example clearly shows the importance of choosing the appropriate formulation because not only the force-displacement curves but also the entire crack growth patterns can be significantly affected for composites that fail by very strongly curved crack paths. We note that shear friction was not considered in the present study, and the crack propagation patterns in experiments can be very different if shear friction plays a significant role.

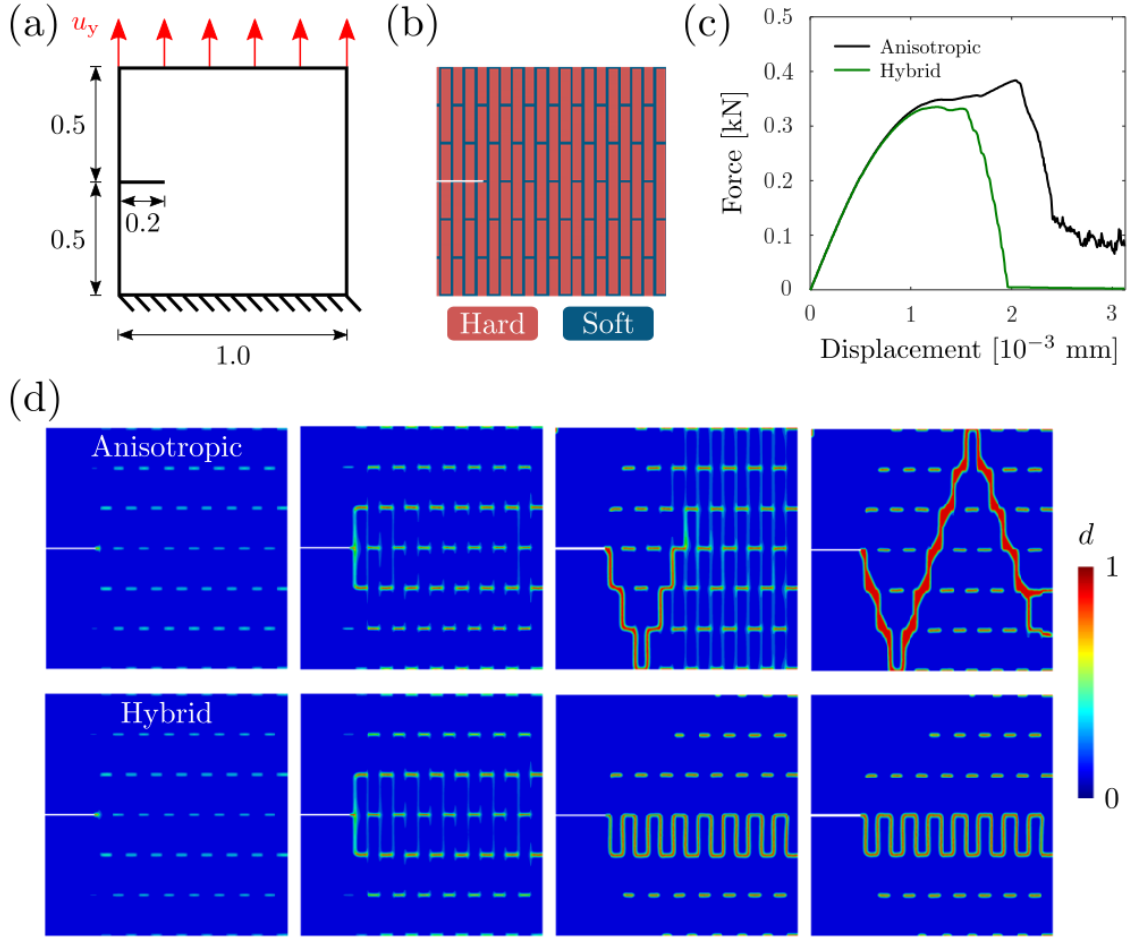

**Fig. 10.** (a) Geometry and boundary condition for single-edge notched specimen (units of measure are mm). (b) Composition of the composite specimen with a nacre-inspired structure (platelets dimensions are 0.04 x 0.320 mm$^2$, the soft interface thickness is 0.012 mm). (c) Force-displacement curves for the tension test. (d) Crack patterns for anisotropic and hybrid formulations at three different loading steps ($u_y = 0.5 \times 10^{-3}$ mm, $u_y = 1.05 \times 10^{-3}$ mm, $u_y = 1.72 \times 10^{-3}$ mm, $u_y = 2.38 \times 10^{-3}$ mm).

## 4. Conclusions

We implemented crack phase field models based on the anisotropic and hybrid formulations, and compared the crack patterns and force-displacement curves in four case studies. The examples include a homogenous specimen fracture under sequential tension and shear loading, a homogeneous beam bending under centered and off-centered indentation, a bone-inspired composite as well as a nacre-inspired composite under pure mode-I loading. For all four cases, the anisotropic formulation was found to overestimate the load because of the remaining finite stiffness of cracks along the shear direction, and it predicts an unphysical crack growth pattern, such as further crack growth after the complete fracture of the specimen and widening of the crack path. Due to complete stiffness degradation of the crack, the hybrid formulation was able to predict both crack patterns and force-displacement curves more naturally than the anisotropic formulation. Our study demonstrates that special care must be taken in the choice of the appropriately coupled strain energy and stiffness degradation schemes to correctly characterize the fracturing of materials under the combination of tensile and shear loading, such as that occurring in heterogeneous biomaterials and bio-inspired composites.

**Acknowledgement**

This research was supported by the Basic Science Research Program (2016R1C1B2011979) and the Creative Materials Discovery Program (2016M3D1A1900038) through the National Research Foundation of Korea (NRF) funded by the Ministry of Science, ICT & Future Planning. This work was also supported by the Korea Institute of Energy Technology Evaluation and Planning (KETEP) and the Ministry of Trade, Industry & Energy (MOTIE) of the Republic of Korea (No. 20174030201480). SS acknowledges the financial support from Ermenegildo Zegna Founder's Scholarship 2017-2018.